\documentclass[amsmath,amssymb,twocolumn]{revtex4}

\usepackage{graphicx,hyperref}
\usepackage[varg]{txfonts}

\newcommand{\ii}{\mathrm{i}}
\newcommand{\e}{\mathrm{e}}
\renewcommand{\d}{\mathrm{d}}

\newcommand{\tens}[1]{\mathbf{#1}}
\newcommand{\cvector}[1]{\left(\begin{array}{c}#1\end{array}\right)}

\begin{document}

\title{Non-equilibrium statistical field theory for classical particles: Linear and mildly non-linear evolution of cosmological density power spectra}
\author{Matthias Bartelmann, Felix Fabis, Daniel Berg, Elena Kozlikin, Robert Lilow, Celia Viermann}
\affiliation{Heidelberg University, Zentrum f\"ur Astronomie, Institut f\"ur Theoretische Astrophysik, Philosophenweg 12, 69120 Heidelberg, Germany}

\begin{abstract}
We use the non-equlibrium statistical field theory for classical particles, recently developed by \citeauthor{2010PhRvE..81f1102M} and \citeauthor{2012JSP...149..643D}, together with the free generating functional we have previously derived for point sets initially correlated in phase space, to calculate the time evolution of power spectra in the free theory, i.e.\ neglecting particle interactions. We provide expressions taking linear and quadratic momentum correlations into account. Up to this point, the expressions are general with respect to the free propagator of the microscopic degrees of freedom.

We then specialise the propagator to that expected for particles in cosmology treated within the Zel'dovich approximation and show that, to linear order in the momentum correlations, the linear growth of the cosmological power spectrum is reproduced. Quadratic momentum correlations return a first contribution to the non-linear evolution of the power spectrum, for which we derive a simple closed expression valid for arbitrary wave numbers. This expression is a convolution of the initial density power spectrum with itself, multiplied by a mode-coupling kernel. We also derive the bispectrum expected in this theory within these approximations and show that its connected part reproduces almost, but not quite, the bispectrum expected in Eulerian perturbation theory of the density contrast.
\end{abstract}

\maketitle

\section{Introduction}

Within the framework of the non-equilibrium field theory for classical particles, developed by \citeauthor{2010PhRvE..81f1102M} and \citeauthor{2012JSP...149..643D} \cite{2010PhRvE..81f1102M, 2012JSP...149..643D}, we have recently derived a free generating functional for canonical, microscopic ensembles of classical particles whose positions and momenta are initially correlated in phase space \cite{paper_1}. While the motivation for such initial ensembles was already derived from cosmological structure-formation theory, this generating functional may be of substantially wider use.

The conventional approaches to the theory of cosmic structure formation perturb the cosmic density field and study the evolution of the perturbations in the Eulerian or Lagrangian pictures (see \cite{1980lssu.book.....P, 2008PhRvD..78h3503B} for reviews). Proceeding into the regime of non-linear density perturbation is notoriously difficult in this approach, and numerous ways have been studied to cope with and to overcome these difficulties, a substantial fraction of which is caused by the dissipation-free nature of the dark-matter particles: Once particle trajectories cross, velocity fields become non-unique, and the density becomes locally arbitrarily large (see \cite{1992MNRAS.254..729B, 1993MNRAS.264..375B, 1994MNRAS.267..811B, 1995A&A...296..575B, 1995ApJ...455....7M, 1997GReGr..29..733E, 2000MNRAS.318..203S, 2001A&A...379....8V, 2002PhR...372....1C, 2004ApJ...612...28M, 2004A&A...421...23V, 2004MNRAS.354.1146V, 2008PhRvD..78h3503B, 2008JCAP...10..036P, 2011JCAP...06..015A, 2012JCAP...12..013A, 2012JHEP...09..082C, 2012JCAP...01..019P, 2013PhRvD..87h3522V, 2014JCAP...07..057C, 2014JCAP...05..022P} for a highly incomplete and admittedly subjective selection of studies).

The non-equilibrium statistical field theory of classical particles avoids these difficulties altogether. It begins with an ensemble of classical particles, whose initial phase-space distribution needs to be suitably correlated in view of cosmological applications. The statistical evolution of this ensemble is fully described by a generating functional. Very much like in statistical quantum field theory, this generating functional consists of operators describing particle interactions, applied to a free generating functional. In its application to classical particle ensembles, operators are switched in between which enable the extraction of collective fields from the free generating functional (see \cite{2010PhRvE..81f1102M, 2011PhRvE..83d1125M, 2013JSP...152..159D, 2012JSP...149..643D} for the pioneering studies).

The free generating functional is fully determined once the initial particle distribution in phase space is specified, and once the free particle propagators are chosen. For linear equations of motion, such as the Hamiltonian equations, the free propagators are easily given.

This approach to cosmic structure formation has two main advantages compared to the conventional approach. First, the density (or any other collective field) is read off the microscopic degrees of freedom in the generating functional when needed, but no dynamical equation for the density needs to be solved. Second, crossing particle trajectories do not affect the theory. Third, even mild perturbations of particle trajectories can lead to substantial perturbations of the density, allowing to intrude deeply into the non-linear regime of density fluctuations, and fourth, the theory naturally contains a damping term which prevents the notorious runaway evolution of non-linear power spectra at large wave numbers.

Here, we use this theory combined with the initially correlated, free generating functional from \cite{paper_1} to derive two- and three-point cumulants of the cosmic particle density. We begin with a brief summary of the theory in Sect.~2. In Sect.~3, we use the formalism to derive closed expressions for the density power spectra, taking momentum correlations into account in linear and quadratic order. As an interlude, we calculate cross-spectra between the density and response fields in Sect.~4.

Cosmological results are derived in Sect.~5, where we first show that the well-known linear evolution of the cosmological power spectrum is recovered in this theory if linear momentum correlations are taken into account, and if the free propagator of the microscopic degrees of freedom is taken from the Zel'dovich approximation \cite{1970A&A.....5...84Z}. For momentum correlations of quadratic order, we obtain a simple and closed expression for a first contribution to the non-linear evolution of the power spectrum, valid for arbitrary wave numbers. Finally, we derive the bispectrum expected in this theory and show that its connected part almost, but not quite reproduces the result from Eulerian perturbation theory of the density field. In Sect.~6, we briefly summarise our results.

\section{Preliminaries}

\subsection{Brief summary of the theory}

We begin by summarising briefly the non-equilibrium field theory for classical particles developed by \citeauthor{2010PhRvE..81f1102M} and \citeauthor{2012JSP...149..643D} \cite{2010PhRvE..81f1102M, 2011PhRvE..83d1125M, 2012JSP...149..643D, 2013JSP...152..159D} together with the generating functional derived in \cite{paper_1} for a canonical ensemble of classical particles whose initial phase-space coordinates are correlated.

A canonical ensemble of $N$ classical point particles can statistically be described by the partition function
\begin{equation}
  Z[H, \tens J, \tens K] = \e^{\ii\hat S_\mathrm{I}}\e^{\ii H\hat\Phi}
  Z_0[\tens J, \tens K]
\label{eq:03-1}
\end{equation}
where $Z_0[\tens J, \tens K]$ is the free generating functional
\begin{equation}
  Z_0[\tens J, \tens K] = \int\d\Gamma_\mathrm{i}\exp\left(
    -\ii\int\left\langle\tens J, \bar{\tens x}\right\rangle\d t
  \right)\;.
\label{eq:03-2}
\end{equation}
Here, $\hat S_\mathrm{I}$ is an operator representing any interaction terms in the particle action. The phase-space trajectory of the free point particle with index $j$, $1\le j\le N$, is given by
\begin{equation}
  \bar x_j(t) = G_\mathrm{R}(t,0)x_j^\mathrm{(i)}-
  \int G_\mathrm{R}(t,t')K(t')\d t'\;,
\label{eq:03-3}
\end{equation}
with the matrix-valued retarded Green's function $G_\mathrm{R}(t,t')$ of the equations of motion for the free particles. The initial phase-space position of the particle $j$ is $x_j^\mathrm{(i)}$. The phase-space coordinates $x_j = (\vec q_j, \vec p_j)$ of all particles are collected in the tensor-valued structure
\begin{equation}
  \tens x = x_j\otimes\vec e_j
\label{eq:03-4}
\end{equation}
with $\vec e_j$ being an $N$-dimensional vector which is unity in its $j$-th component and zero elsewhere. Likewise, we introduce
\begin{equation}
  \mathcal{G}(t,t') = G_\mathrm{R}(t,t')\otimes\vec e_j
\label{eq:03-5}
\end{equation}
to collect the Green's functions acting on all particles in one object. Finally, the source fields
\begin{equation}
  \tens J = \cvector{\vec J_q\\ \vec J_p}\otimes\vec e_j\;,\quad
  \tens K = \cvector{\vec K_q\\ \vec K_p}\otimes\vec e_j
\label{eq:03-6}
\end{equation}
similarly collect the source-field components for the positions and momenta of all $N$ particles. The angular brackets denote the scalar product
\begin{equation}
  \left\langle\tens x,\tens y\right\rangle :=
  \left(x_j\otimes\vec e_j\right)\cdot\left(y_k\otimes\vec e_k\right) =
  \left(x_j\cdot y_k\right)\otimes\left(\vec e_j\cdot\vec e_k\right) =
  x_j\cdot y_j\;,
\label{eq:03-7}
\end{equation}
where summation over repeated indices is implied.

Inserting
\begin{equation}
  \bar{\tens x}(t) = \mathcal{G}(t,0)\tens x^\mathrm{(i)}-
  \int\mathcal{G}(t,t')\tens K(t')\d t'
\label{eq:03-8}
\end{equation}
into (\ref{eq:03-2}), we can factorise the free generating functional into a factor $Z_0$ depending on the initial phase-space positions $\tens x^\mathrm{(i)}$ and a factor depending on the sources $\tens J$ and $\tens K$ only,
\begin{equation}
  Z_0[\tens J, \tens K] = \e^{-\ii S_K[\tens J, \tens K]}\bar Z_0[\tens J]
\label{eq:03-9}
\end{equation}
with
\begin{equation}
  \bar Z_0[\tens J] = \int\d\Gamma_\mathrm{i}\exp\left(
    -\ii\int\left\langle
      \tens J, \mathcal{G}(t,0)\bar{\tens x}^\mathrm{(i)}
    \right\rangle\d t
  \right)
\label{eq:03-10}
\end{equation}
and
\begin{equation}
  S_K[\tens J, \tens K] = \int\d t\int\d t'\left\langle
    \tens J(t),\mathcal{G}(t,t')\tens K(t')
  \right\rangle\;.
\label{eq:03-11}
\end{equation}
Since the initial phase-space positions $\tens x^\mathrm{(i)}$ can be pulled out of the time integral in (\ref{eq:03-10}), we introduce the time-integrated source vectors $\bar{\tens J}_q$ and $\bar{\tens J}_p$ by
\begin{equation}
  \bar{\tens J}_{q,p}^\top := \int
    \left\langle\tens J(t),\mathcal{G}(t,0)\mathcal{P}_{q,p}\right\rangle
  \d t\;,
\label{eq:03-12}
\end{equation}
defined by means of the projectors
\begin{equation}
  \mathcal{P}_q := \cvector{\mathcal{I}_3\\ 0_3}\otimes\mathcal{I}_N\;,\quad
  \mathcal{P}_p := \cvector{0_3\\ \mathcal{I}_3}\otimes\mathcal{I}_N\;.
\label{eq:03-13}
\end{equation} 
Here, $\mathcal{I}_3$ is the unit matrix in three dimensions, and $0_3$ is a $3\times3$ matrix whose elements are all zero. In terms of $\bar{\tens J}_{q,p}$,
\begin{equation}
  \bar Z_0[\tens J] = \int\d\Gamma_\mathrm{i}\e^{
    \ii\left\langle\bar{\tens J}_q,\tens q\right\rangle+
    \ii\left\langle\bar{\tens J}_q,\tens p\right\rangle
  }\;.
\label{eq:03-14}
\end{equation} 

The integral measure in phase space,
\begin{equation}
  \d\Gamma_\mathrm{i} = P\left(\tens q,\tens p\right)\d\tens q\d\tens p\;,
\label{eq:03-15}
\end{equation}
weighs the individual phase-space points with an initial probability distribution $P(\tens q, \tens p)$. Below, we shall specify $P(\tens q, \tens p)$ to be the phase-space probability distribution
\begin{equation}
  P(\tens q, \tens p) =
  \frac{A}{\sqrt{(2\pi)^{3N}\det C_{pp}}}\,\mathcal{C}(\tens p)
  \exp\left(-\frac{1}{2}\tens p^\top C_{pp}^{-1}\tens p\right)\;,
\label{eq:03-16}
\end{equation}
derived in \cite{paper_1} for a point set whose initial correlations are expressed by the correlation operator $\mathcal{C}(\tens p)$ containing correlations between spatial positions as well as between spatial positions and momenta, and by the momentum correlation matrix $C_{pp}$.

\subsection{Density cumulants}

Based on the free generating functional $Z_0$ for the microscopic degrees of freedom, all required collective fields are expressed by the operator tuple $\hat\Phi$. At least, $\hat\Phi$ has two components, a density operator $\hat\Phi_\rho$ and an operator $\hat\Phi_B$ for the response field required for describing the interaction between particles. In Fourier space, the operator for the contribution of particle $j$ to the density, acting at the arbitrary position $1 := (t_1, \vec q_1)$, is
\begin{equation}
  \hat\Phi_{\rho_j}(1) = \exp\left(
    -\vec k_1\cdot\frac{\delta}{\delta\vec J_{q_j}(1)}
  \right)\;,
\label{eq:03-17}
\end{equation}
where $\vec k_1$ is the Fourier wave vector conjugate to the position $\vec q_1$. The operator for the complete density is the sum over the one-particle density operators,
\begin{equation}
  \hat\Phi_\rho(1) = \sum_{j=1}^N\hat\Phi_{\rho_j}(1)\;.
\label{eq:03-18}
\end{equation}
When applied to the free generating functional $Z_0$, the one-particle density operator simply translates the source $\tens J$,
\begin{equation}
  \hat\Phi_{\rho_j}(1)Z_0[\tens J, \tens K] =
  Z_0[\tens J+\tens L_j(1), \tens K]\;,
\label{eq:03-19}
\end{equation}
by the amount
\begin{equation}
  \tens L_j(1) = -\delta_\mathrm{D}\left(t-t_1\right)
  \cvector{\vec k_1 \\ 0}\otimes\vec e_j\;.
\label{eq:03-20}
\end{equation}

The application of further density operators leads to further translations. The result of applying any sequence of an arbitrary number $m$ of one-particle density operators will thus lead to
\begin{equation}
  \hat\Phi_{\rho_{j_1}}(1)\ldots\hat\Phi_{\rho_{j_m}}(m)Z_0[\tens J,\tens K] =
  Z_0[\tens J+\tens L,\tens K]
\label{eq:03-21}
\end{equation}
with
\begin{equation}
  \tens L(t) := -\sum_{s=1}^m\delta_\mathrm{D}(t-t_s)\cvector{\vec k_s \\ 
0}\otimes\vec e_{j_s}\;.
\label{eq:03-22}
\end{equation}

Once all necessary density operators have been applied, the source $\tens J$ can be switched off. Writing $Z[\tens J, \tens K]$ in the representation (\ref{eq:03-9}), we can then write
\begin{equation}
  \left.Z_0[\tens J+\tens L,\tens K]\right\vert_{\tens J=0} = \e^{-\ii 
S_K[\tens L,\tens K]}\bar Z_0[\tens L]\;.
\label{eq:03-23}
\end{equation}
In other words, the application of an arbitrary sequence of density operators results in replacing $\tens J$ by $\tens L$ in the phase-space integrated generating functional $\bar Z_0[\tens J]$ given in (\ref{eq:03-14}). The time-integrated source $\bar{\tens J}$ then needs to be replaced by the time-integrated shift tensor $\bar{\tens L}$ with the components
\begin{equation}
  \bar{\tens L}_{q,p}^\top = \int
    \left\langle\tens L(t),\mathcal{G}(t,0)\mathcal{P}_{q,p}\right\rangle
  \d t\;.
\label{eq:03-24}
\end{equation}
Inserting the shift (\ref{eq:03-22}) into these equations, we find
\begin{equation}
  \bar{\tens L}_q = -\sum_{s=1}^m\vec k_s\otimes\vec e_{j_s}\;,\quad
  \bar{\tens L}_p = -\sum_{s=1}^mg_{qp}(t_s,0)\vec k_s\otimes\vec e_{j_s}\;.
\label{eq:03-25}
\end{equation}

According to (\ref{eq:03-21}), an $m$-point density cumulant $G_{\rho\ldots\rho}(1\ldots m)$ is found by summing over all particle indices,
\begin{equation}
  G_{\rho\ldots\rho}(1\ldots m) = \sum_{j_1\ldots j_m=1}^N
  \hat\Phi_{\rho_{j_1}}\cdots\hat\Phi_{\rho_{j_m}}\bar Z_0[\tens J] =
  \sum_{j_1\ldots j_m=1}^N\bar Z_0[\tens L]\;.
\label{eq:03-26}
\end{equation}
The further procedure now consists in specifying the shift tensor $\tens L$ and its components, and to insert the results in place of the source components $\bar J_{q_j}$ and $\bar J_{p_j}$ into the expressions derived in \cite{paper_1} for the contributions to the free generating functional. Specifically, we have
\begin{equation}
  \bar L_{q_j} = \left\langle
    \bar{\tens L}_q,\mathcal{I}_3\otimes\vec e_j
  \right\rangle =
  -\sum_{s=1}^m\vec k_s\delta_{jj_s}
\label{eq:03-27}
\end{equation}
and likewise
\begin{equation}
  \bar L_{p_j} = -\sum_{s=1}^mg_{qp}(t_s,0)\vec k_s\delta_{jj_s}\;.
\label{eq:03-28}
\end{equation}
Thus, the time-integrated shift vectors $\bar{\tens L}_{q,p}$ have non-vanishing components only where the component index agrees with the particle index $j_s$ set by the one-particle density operator applied to the free generating functional. For any shift tensor specified by a complete set of $m$ particle indices $j_1\ldots j_m$, we write
\begin{equation}
  \bar Z_0[\tens L] = G_{j_1\ldots j_m}\;,\quad
  G_{\rho\ldots\rho}(1\ldots m) = \sum_{j_1\ldots j_m=1}^NG_{j_1\ldots j_m}\;.
\label{eq:03-29}
\end{equation}

\subsection{Contributions from linear and quadratic momentum correlations}

We now proceed to work out the terms $G_{j_1\ldots j_m}$ for $m=2$ and $m=3$. For doing so, we need the result
\begin{equation}
  \bar Z_{jk}^{(1)} = 
  (2\pi)^3\delta_\mathrm{D}\left(\bar L_{q_j}+\bar L_{q_k}\right)
  \mathcal{N}'_{jk}
  P_\delta\left(\bar L_{q_j}\right)A_{jk}^2\left(\bar L_{q_j}\right)
\label{eq:03-30}
\end{equation}
from \cite{paper_1} for the terms contributed by linear momentum correlations to the free generating functional. Here, the abbreviations
\begin{equation}
  \mathcal{N}'_{jk} := \int\d\tens q'\,
  \e^{\ii\left\langle\bar{\tens L}_q,\tens q'\right\rangle}
\label{eq:03-31}
\end{equation}
and
\begin{equation}
  A_{jk}^2\left(\bar L_{q_j}\right) := \frac{1}{2}\left(
    1-a_{jk}^2\left(\bar L_{q_j}\right)
  \right)-b_{jk}\left(\bar L_{q_j}\right)\;,
\label{eq:03-32}
\end{equation}
containing
\begin{equation}
  a_{jk}^2\left(\bar L_{q_j}\right) :=
  \frac{\left(\bar L_{p_j}\cdot\bar L_{q_j}\right)
        \left(\bar L_{q_j}\cdot\bar L_{p_k}\right)}
  {\bar L_{q_j}^{\,4}} \;,\quad
  b_{jk}\left(\bar L_{q_j}\right) :=
  \frac{\bar L_{q_j}\cdot\bar L_{p_k}}{\bar L_{q_j}^{\,2}}
\label{eq:03-33}
\end{equation}
were used. The prime in (\ref{eq:03-31}) indicates that $\vec q_j$ and $\vec q_k$ are to be excluded from the integration over all spatial particle positions $\tens q$.

Of the terms contributed by quadratic momentum correlations, we only need the expressions
\begin{align}
  &\bar Z_{jkjk}^{(2)} = (2\pi)^3
  \delta_\mathrm{D}\left(\bar L_{q_j}+\bar L_{q_k}\right)\mathcal{N}'_{jk}
  \nonumber\\ &\times
  \int\frac{\d^3k}{(2\pi)^3}
  P_\delta\left(\vec k\,\right)P_\delta\left(\vec k-\bar L_{q_j}\right)
  a_{jk}^2\left(\vec k\,\right)a_{jk}^2\left(\vec k-\bar L_{q_j}\right)
\label{eq:03-34}
\end{align}
and
\begin{align}
  &\bar Z_{jkkl}^{(2)} = (2\pi)^3
  \delta_\mathrm{D}\left(\bar L_{q_j}+\bar L_{q_k}+\bar L_{q_l}\right)
  \mathcal{N}'_{jkl} \nonumber\\ &\times
  P_\delta\left(\bar L_{q_j}\right)P_\delta\left(\bar L_{q_l}\right)
  a_{jk}^2\left(\bar L_{q_j}\right)a_{kl}^2\left(\bar L_{q_l}\right)\;.
\label{eq:03-35}
\end{align}
With those expressions, the generating functional is
\begin{equation}
  \bar Z_0[\tens L] \approx \bar Z_0^{(1)}[\tens L]+\bar Z_0^{(2)}[\tens L]
\label{eq:03-36}
\end{equation}
with
\begin{align}
  \bar Z_0^{(1)}[\tens L] &= V^{-N}\e^{-Q_D/2}\sum_{j\ne k=1}^N
  \bar Z_{jk}^{(1)} \;,\nonumber\\
  \bar Z_0^{(2)}[\tens L] &= \frac{V^{-N}}{8}\e^{-Q_D/2}\sum_{j\ne k, l\ne m=1}^N
  \bar Z_{jklm}^{(2)}\;,
\label{eq:03-37}
\end{align}
where the damping term
\begin{equation}
  Q_D = \frac{\sigma_1^2}{3}\sum_{j=1}^N\bar L_{p_j}^{\,2}
\label{eq:03-38}
\end{equation}
appears.

\section{Two-point density cumulants}

\subsection{Linear momentum correlations}

Since the particles cannot be distinguished, it suffices to select any set of $m$ out of the $N$ particles to evaluate the remaining sum in (\ref{eq:03-29}). These $m$ particles can be labelled with indices from $1$ to $m$ without loss of generality. The generating functional $\bar Z_0[\tens L]$ then needs to be calculated for this specific selection of particles, and the resulting terms multiplied by the number of possibilities for the particular subset of $m$ particles selected from the canonical ensemble of $N$ particles.

If we wish to calculate density cumulants taking momentum correlations into account to first or second order, we need to evaluate the expressions $\bar Z_{jk}^{(1)}$ from Eq.~(128) and $\bar Z_{jklm}^{(2)}$ given in Eqs.~(132), (133) and (134) of \cite{paper_1}.

For a two-point cumulant, $m = 2$, we can choose $j_1,j_2 \in \{1,2\}$. Since the particles have to be different for the correlation terms in $\bar Z_{jk}^{(1)}$ and $\bar Z_{jklm}^{(2)}$ not to vanish, we set $(j_1,j_2) = (1,2)$. Then,
\begin{equation}
  \bar L_{q_1} = -\vec k_1\;,\quad\bar L_{q_2} = -\vec k_2\;,
\label{eq:03-39}
\end{equation}
accordingly
\begin{equation}
  \bar L_{p_1} = -g_{qp}(t_1,0)\vec k_1\;,\quad
  \bar L_{q_2} = -g_{qp}(t_2,0)\vec k_2\;,
\label{eq:03-40}
\end{equation}
and no other components of $\bar{\tens L}_{q,p}$ appear. Then, the remaining integrals over all positions except $\vec q_1$ and $\vec q_2$ simply give
\begin{equation}
  \mathcal{N}_{12}' = V^{N-2}\;.
\label{eq:03-41}
\end{equation}
If we set $t_1 = t_2$, i.e.\ if the cumulant is taken synchronously, we further have
\begin{align}
  A_{12}^2\left(\vec k_1\right) &= \frac{1}{2}\left(
    1+g_{qp}^2(t_1,0)
  \right)+g_{qp}(t_1,0) \nonumber\\ &=
  \frac{1}{2}\left(1+g_{qp}(t_1,0)\right)^2\;,
\label{eq:03-42}
\end{align}
taking into account that the remaining delta distribution ensures $\vec k_1 = -\vec k_2$. Therefore,
\begin{equation}
  \bar Z_{12}^{(1)} = \frac{V^{N-2}}{2}
  (2\pi)^3\delta_\mathrm{D}\left(\vec k_1+\vec k_2\right)
  \left(1+g_{qp}(t_1,0)\right)^2P_\delta\left(\vec k_1\right)\;.
\label{eq:03-43}
\end{equation}
This expression is symmetric under the permutation $(1,2)\to(2,1)$. Since the index pair $(1,2)$ can be selected in $N(N-1)\approx N^2$ ways from the $N$ particles, we immediately find
\begin{align}
  G_{\rho\rho}^{(1)}(12) &= \e^{-Q_D/2}\bar\rho^2
  (2\pi)^3\delta_\mathrm{D}\left(\vec k_1+\vec k_2\right)
  \nonumber\\ &\times
  \left(1+g_{qp}(t_1,0)\right)^2P_\delta\left(\vec k_1\right)\;,
\label{eq:03-44}
\end{align}
with the damping term
\begin{equation}
  Q_D = \frac{\sigma_1^2}{3}g_{qp}^2(t_1,0)k_1^2\;,
\label{eq:03-45}
\end{equation}
where $\vec k_2 = -\vec k_1$ was used once more.

\subsection{Quadratic momentum correlations}

Proceeding to the contribution of quadratic momentum correlations to the two-point cumulant, we see immediately that only terms of the form
\begin{align}
  &\bar Z_{jkjk}^{(2)} = (2\pi)^3
  \delta_\mathrm{D}\left(\bar J_{q_j}+\bar J_{q_k}\right)\mathcal{N}'_{jk}
  \nonumber\\ &\times
  \int\frac{\d^3k}{(2\pi)^3}
  P_\delta\left(\vec k\,\right)P_\delta\left(\vec k-\bar J_{q_j}\right)
  a_{jk}^2\left(\vec k\,\right)a_{jk}^2\left(\vec k-\bar J_{q_j}\right)\;,
\label{eq:03-46}
\end{align}
derived in \cite{paper_1} can contribute because terms with three or four different particle indices must vanish for a two-point cumulant. Setting again $(j_1,j_2)=(1,2)$, and evaluating the factors $a_{jk}^2 = a_{12}^2$ with the appropriate momentum shift vectors (\ref{eq:03-40}), we immediately arrive at
\begin{align}
  &\bar Z_{1212}^{(2)} = V^{N-2}
  (2\pi)^3\delta_\mathrm{D}\left(\vec k_1+\vec k_2\right)
  g_{qp}^4(t_1,0) \nonumber\\ &\times
  \int\frac{\d^3k}{(2\pi)^3}
  P_\delta\left(\vec k\,\right)P_\delta\left(\vec k-\vec k_1\right)
  \left(\frac{\vec k_1\cdot\vec k}{k^2}\right)^2
  \left(
    \frac{\vec k_1\cdot(\vec k-\vec k_1)}{(\vec k-\vec k_1)^2}
  \right)^2\;.
\label{eq:03-47}
\end{align}

As derived in \cite{paper_1}, the terms $\bar Z_{jklm}^{(2)}$ are symmetric in the first and second index pair and under exchanges of the two index pairs, and the indices in the first and second index pairs must be different. Under these requirements, the term (\ref{eq:03-47}) appears $4$ times in the sum over particle indices: terms with the index combinations $(1212)$, $(2112)$, $(1221)$ and $(2121)$ are all equivalent, and others do not appear. Furthermore, we have to multiply with the number $N(N-1) \approx N^2$ of ways for selecting a pair from the $N$ particles. Thus, we arrive at the contribution
\begin{align}
  &G_{\rho\rho}^{(2)}(12) = \e^{-Q_D/2}\frac{\bar\rho^2}{2}
  (2\pi)^3\delta_\mathrm{D}\left(\vec k_1+\vec k_2\right)
  g_{qp}^4(t_1,0) \nonumber\\ &
  \times\int\frac{\d^3k}{(2\pi)^3}
  P_\delta\left(\vec k\,\right)P_\delta\left(\vec k-\vec k_1\right)
  \left(\frac{\vec k_1\cdot\vec k}{k^2}\right)^2
  \left(
    \frac{\vec k_1\cdot(\vec k-\vec k_1)}{(\vec k-\vec k_1)^2}
  \right)^2
\label{eq:03-48}
\end{align}
for the contribution of quadratic momentum correlations to the two-point density cumulant. The damping term $Q_D$ equals that given in (\ref{eq:03-45}).

\section{Response-field cumulants}

\subsection{General expressions and mean response field}

It was shown in \cite{paper_1} that the response field $B$ can be obtained in Fourier space by applying an operator $\hat\Phi_{B}$ to the free generating functional whose one-particle representation can be expressed by a one-particle density operator,
\begin{equation}
  \hat\Phi_{B_j}(1) = \left(
    \vec k_1^{\,\top}\cdot \frac{\delta}{\delta\vec K_{p_j}(1)}
  \right)\hat\Phi_{\rho_j}(1) =:
  \hat b_j(1)\hat\Phi_{\rho_j}(1)\;,
\label{eq:03-49}
\end{equation}
multiplied by the factor
\begin{equation}
  b_j(1) := -\ii\vec k_1^{\,\top}\cdot
  \frac{\delta S_K[\tens L_j(1),\tens K]}{\delta\vec K_{p_j}(1)}\;.
\label{eq:03-50}
\end{equation}

This expression, valid for a single response-field operator applied to the free generating functional, is easily generalised. Suppose we apply $m$ operators in total, of which $n$ are density and $m-n$ are response-field operators. Since each response-field operator contains a density operator to be executed first, we will have to apply $m$ density operators in total. The result will have to be multiplied by $m-n$ response-field factors. Thus, we have
\begin{align}
  &\underbrace{\hat\Phi_B(m)\ldots\hat\Phi_B(m-n)}_{m-n\;\mathrm{terms}}
  \underbrace{\hat\Phi_\rho(n)\ldots\hat\Phi_\rho(1)}_{n\;\mathrm{terms}}\,
  Z_0[\tens J, \tens K] \nonumber\\
  &= b(m)\ldots b(m-n)\hat\Phi_\rho(m)\ldots\hat\Phi_\rho(1)\,
  Z_0[\tens J, \tens K] \nonumber\\
  &= b(m)\ldots b(m-n)\,Z_0[\tens J+\tens L,\tens K]\;.
\label{eq:03-51}
\end{align}

Decomposing the density operators and the response-field into their single-particle contributions, we obtain the shift tensor $\tens L$ from (\ref{eq:03-22}). After inserting this into $S_K$ from (\ref{eq:03-11}), any single-particle response-field operator $\hat b_{j_l}(l)$ returns the factor
\begin{equation}
  b_{j_l}(l) = \ii\sum_{s=1}^mg_{qp}(t_s,t_l)\,\vec k_s\cdot\vec k_l\,
  \delta_{j_lj_s}\;.
\label{eq:03-52}
\end{equation}

Applying a single response-field operator to the generating functional, we obtain the mean response field. In the general approach sketched above, we set $m = 1 = n$. Then, from (\ref{eq:03-52}), we have
\begin{equation}
  b_{j_1}(1) = \ii g_{qp}(t_1, t_1)\,k_1^2 = 0
\label{eq:03-53}
\end{equation}
if the propagator $g_{qp}(t,t')$ vanishes for $t = t'$, as it will usually do. Then, the mean response field vanishes identically.

\subsection{Response-field power spectra}

For $m = 2$, we have the density-response cumulant
\begin{align}
  G_{\rho_{j_2}B_{j_1}}(12) &=
  \ii\vec k_1\cdot\left(
    g_{qp}(t_1,t_1)\vec k_1\delta_{j_1j_1}+
    g_{qp}(t_1,t_2)\vec k_2\delta_{j_1j_2}
  \right) \nonumber\\ &\times G_{\rho_{j_1}\rho_{j_2}}(12)
\label{eq:03-54}
\end{align}
according to (\ref{eq:03-52}). Since $g_{qp}(t_1,t_1) = 0$, the first term in parentheses vanishes. The second term contributes only if $j_1 = j_2$ because of the Kronecker symbol, but then the correlation term in (\ref{eq:03-29}) cannot contribute. Since
\begin{equation}
  \mathcal{N} = (2\pi)^3\delta_\mathrm{D}\left(\vec k_1+\vec k_2\right)V^{N-1}
\label{eq:03-55}
\end{equation}
in this case and
\begin{align}
  Q_D &= \frac{\sigma_1^2}{3}\left(
    g_{qp}(t_1,0)\vec k_1+g_{qp}(t_2,0)\vec k_2
  \right)^2 \nonumber\\ &=
  \frac{\sigma_1^2}{3}\left(g_{qp}(t_1,0)-g_{qp}(t_2,0)\right)^2k_1^2\;,
\label{eq:03-56}
\end{align}
taking $\vec k_2=-\vec k_1$ into account, we find
\begin{align}
  G_{\rho B}(12) &= -\ii\bar\rho(2\pi)^3
  \delta_\mathrm{D}\left(\vec k_1+\vec k_2\right)\,g_{qp}(t_1,t_2)k_1^2
  \nonumber\\ &\times
  \exp\left(
    -\frac{\sigma_1^2k_1^2}{6}
    \left(g_{qp}(t_1,0)-g_{qp}(t_2, 0)\right)^2
  \right)
\label{eq:03-57}
\end{align}
after summing over all $1\le j_1\le N$. Changing the order of $B$ and $\rho$ in (\ref{eq:03-54}) only changes the ordering of the times $t_1$ and $t_2$, thus leading to
\begin{align}
  G_{B\rho}(12) &= -\ii\bar\rho(2\pi)^3
  \delta_\mathrm{D}\left(\vec k_1+\vec k_2\right)\,g_{qp}(t_2,t_1)k_1^2
  \nonumber\\ &\times
  \exp\left(
    -\frac{\sigma_1^2k_1^2}{6}
    \left(g_{qp}(t_1,0)-g_{qp}(t_2, 0)\right)^2
  \right)\;.
\label{eq:03-58}
\end{align}
The cross-spectra $G_{\rho B}(12)$ and $G_{B\rho}(12)$ will obviously vanish if $t_1 = t_2$. Applying both $\hat b_{j_1}(1)$ and $\hat b_{j_2}(2)$ will return a product of propagators with time orderings $(t_1,t_2)$ and $(t_2,t_1)$, which must vanish for causality, hence
\begin{equation}
  G_{BB}(12) = 0\;.
\label{eq:03-59}
\end{equation} 

\section{Cosmological power spectra}

\subsection{Hamiltonian vs.\ Zel'dovich propagators}

It was shown in \cite{paper_2} that a free classical point particle in an expanding space-time has the propagator
\begin{equation}
  g_{qp}(\tau,\tau') = \int_{\tau'}^\tau\frac{\d\bar\tau}{g(\bar\tau)}\;,
\label{eq:03-60}
\end{equation}
where the growth factor $D_+(a)$ is used as the time coordinate $\tau$,
\begin{equation}
  \tau = D_+(a)-1\;,
\label{eq:03-61}
\end{equation}
set to zero at some early time. The scale factor $a$ of the expanding space-time is set to unity at the initial epoch, $a_\mathrm{i} = 1$, and the growth factor is normalised such as to equal unity at $a_\mathrm{i}$. The function $g(\tau)$ in (\ref{eq:03-60}) is
\begin{equation}
  g(\tau) = a^2D_+(a)f\frac{H}{H_\mathrm{i}}\;,
\label{eq:03-62}
\end{equation}
with the Hubble function $H$, its value $H_\mathrm{i}$ at the initial time, and
\begin{equation}
  f := \frac{\d\ln D_+}{\d\ln a}\;.
\label{eq:03-63}
\end{equation}

In an Einstein-de Sitter universe, $D_+(a) = a$, thus $f = 1$, and $HH^{-1}_\mathrm{i} = a^{-3/2}$. Therefore, $\tau = a-1$, hence $g(\tau) = (1+\tau)^{3/2}$ and
\begin{equation}
  g_{qp}(\tau,\tau') = 2\left(
    \frac{1}{\sqrt{1+\tau'}}-\frac{1}{\sqrt{1+\tau}}
  \right) \le 2\;.
\label{eq:03-64}
\end{equation}
Even though the finite bound on $g_{qp}(\tau,\tau')$ was derived for an Einstein-de Sitter model here, it remains true in general cosmological models at least in the matter-dominated phase that the propagator $g_{qp}(\tau,\tau')$ remains finite even for $\tau' = 0$ and $\tau\to\infty$.

Clearly, therefore, inserting the free Hamiltonian propagator (\ref{eq:03-60}) into the free linear power-spectrum term (\ref{eq:03-60}) cannot reproduce the result well-known from ordinary cosmological perturbation theory that the matter power spectrum evolves linearly as $P_\delta(k) \propto D_+^2(a)$.

We can, however, easily achieve this behaviour if we replace the Hamiltonian propagator (\ref{eq:03-60}) by the Zel'dovich propagator
\begin{equation}
  g_{qp}^\mathrm{(Z)}(\tau,\tau') = \tau-\tau'\;.
\label{eq:03-65}
\end{equation}
Then, the time-evolution factor in (\ref{eq:03-60}) turns into
\begin{equation}
  1+g_{qp}(t_1,0) \to 1+\tau_1 = D_+(a)\;,
\label{eq:03-66}
\end{equation}
and the free linear power-spectrum contribution (\ref{eq:03-60}) scales as $P_\delta(k) \propto D_+^2(a)$ as it should.

This replacement of the free Hamiltonian by the Zel'dovich propagator is justified and advantageous because the Zel'dovich approximation already contains part of the gravitational interaction between the particles. With the free Hamiltonian propagator, the power spectra shown here would develop without any gravitational interaction taken into account. Instead, by the gravitational interaction it already includes, the Zel'dovich propagator reproduces the linear growth of the density power spectrum at the lowest order, i.e.\ with the free generating functional and the momentum correlations taken into account at linear order only.

With the Zel'dovich propagator, the free contribution (\ref{eq:03-47}) to the power spectrum with quadratic momentum correlations is proportional to the fourth power of $\tau = D_+-1$, i.e.\ both convolved power spectra evolve linearly.

\subsection{Damping}

The damping term
\begin{equation}
  \exp\left(-\frac{\sigma_1^2}{3}g_{qp}^2(t_1,0)k_1^2\right)
\label{eq:03-67}
\end{equation} 
present in both contributions (\ref{eq:03-44}) and (\ref{eq:03-48}) requires a separate consideration. It arises from the streaming of the particles with the initial root mean-square velocity quantified by $\sigma_1$. In cold dark matter, free streaming is suppressed by the gravitational interaction between the particles. Since gravitational interaction is not included yet beyond the part contained in the Zel'dovich approximation at the level of the theory presented here, the damping is both inevitable and unrealistic. We take account of this fact by ignoring the damping term completely at the lowest order of the theory, i.e.\ when considering the free theory with the Zel'dovich propagator and the linear momentum correlations.

At the next higher order of the free theory, i.e.\ when the momentum correlations are taken into account quadratically, the damping term needs to be switched on in an appropriate way. If we could include the complete momentum-correlation hierarchy, the complete exponential damping term from (\ref{eq:03-67}) would also have be included. However, since the quadratic momentum correlation goes only one step beyond the lowest order, it is appropriate to approximate the factor (\ref{eq:03-67}) also at lowest non-trivial order. We write
\begin{equation}
  \exp\left(-\frac{\sigma_1^2}{3}g_{qp}^2(t_1,0)k_1^2\right) \approx
  \frac{1}{1+\frac{\sigma_1^2}{3}g_{qp}^2(t_1,0)k_1^2}
\label{eq:03-68}
\end{equation}
and multiply the power-spectrum contribution (\ref{eq:03-47}) by this approximate damping factor instead of the complete exponential.

\subsection{Power-spectra contributions from the free generating functional}

\begin{figure}
  \includegraphics[width=\hsize]{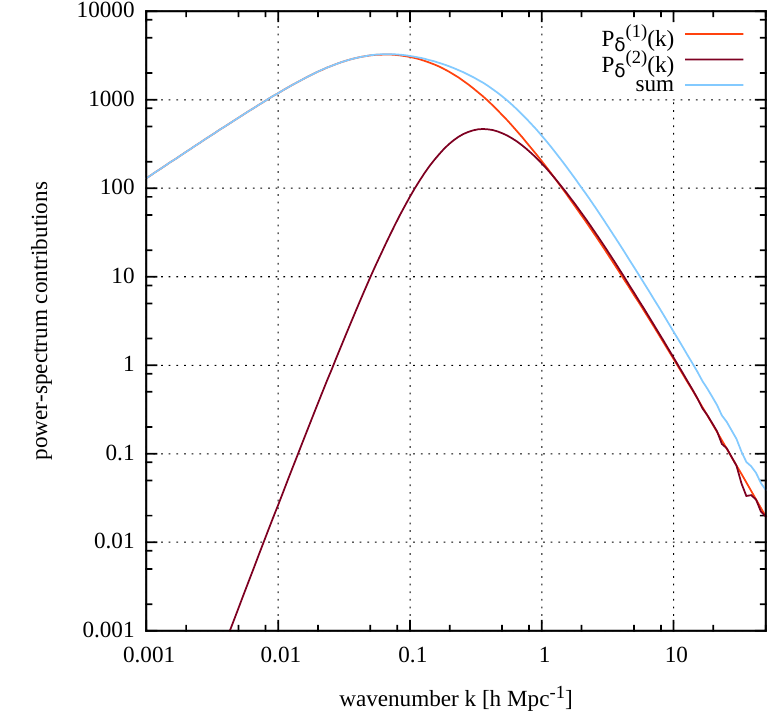}
\caption{The contributions $P_\delta^{(1)}(k)$ and $P_\delta^{(2)}(k)$ to the density power spectrum obtained from the free generating functional with the Zel'dovich propagator are shown for a CDM initial power spectrum, evolved to $\sigma_8 = 0.8$. The fluctuations in the term $P_\delta^{(2)}(k)$ at the large-wavenumber end are numerical noise from a Monte-Carlo integration.}
\label{fig:1}
\end{figure}

We thus obtain the power-spectrum contributions in the free theory with the Zel'dovich propagator and the suitably approximated damping terms
\begin{align}
  &P_\delta^{(1)}(k) = P_\delta^\mathrm{(i)}\left(k\right)
  \left(1+\tau_1\right)^2 = D_+^2P_\delta^\mathrm{(i)}(k)\;,\nonumber\\
  &P_\delta^{(2)}(k) = \frac{\tau_1^4}{2\left(
    1+\frac{\sigma_1^2}{3}\tau_1^2k^2
  \right)} \nonumber\\ &\times
  \int\frac{\d^3k'}{(2\pi)^3}
  P_\delta^\mathrm{(i)}\left(k'\right)
  P_\delta^\mathrm{(i)}\left(\vec k-\vec k'\right)
  \left(\frac{\vec k\cdot\vec k'}{k'^2}\right)^2
  \left(
    \frac{\vec k\cdot(\vec k-\vec k')}{(\vec k-\vec k')^2}
  \right)^2\;,
\label{eq:03-69}
\end{align}
where we have specified for clarity that the power spectra on the right-hand sides are the power spectra characterising the initial particle distribution. The two contributions (\ref{eq:03-69}) are shown in Fig.~\ref{fig:1} for an initial CDM power spectrum evolved in an Einstein-de Sitter universe such that its present normalisation reaches $\sigma_8 = 0.8$. The fluctuations visible there in the $P_\delta^{(2)}(k)$ term are numerical noise from a Monte-Carlo integration of the convolution remaining in (\ref{eq:03-69}). With the Monte-Carlo integrator contained in the Gnu Science Library, evaluating the integrand in (\ref{eq:03-69}) $10^5$ times, the curves in Fig.~\ref{fig:1} require $\sim30$~seconds on a single core of a slightly outdated desktop PC.

\subsection{Bispectrum}

The contribution to the bispectrum due to linear momentum correlations can be easily read off (\ref{eq:03-29}). Ignoring the damping term and focussing on the correlated contribution to the free generating functional $\bar Z_0^{(1)}[\tens L]$ in (\ref{eq:03-29}), we first notice that again neither $\bar L_{q_1}$ nor $\bar L_{q_2}$ must vanish since the result would otherwise be zero. Therefore, at least one each of the particle indices $(j_1,j_2,j_3)$ must be set to $1$ and $2$, while the third particle index available for the bispectrum remains free.

If we set this third index to $1$ or $2$ as well, the multiplicity of the resulting term is $\propto N^2$, which is lower by a factor of $N$ than the multiplicity $\propto N^3$ required for the bispectrum. This term is thus negligible. Only terms with the third index set to $>2$ will remain. Adoping $(j_1,j_2,j_3) = (1,2,3)$ implies
\begin{equation}
  \bar L_{q_1} = -\vec k_1\;,\quad
  \bar L_{q_2} = -\vec k_2\;,\quad
  \mathcal{N}_{12}' = (2\pi)^3\delta_\mathrm{D}\left(\vec k_3\right)V^{N-3}\;.
\label{eq:03-70}
\end{equation}
Moreover, for a synchronous bispectrum, $\tau_1 = \tau_2 = \tau_3$. Using the Zel'dovich propagator,
\begin{equation}
  A_{12}^2 = \frac{1}{2}\left(
    1-\tau_1^2\frac{\vec k_1\cdot\vec k_2}{k_1^2}
  \right)-\tau_1\frac{\vec k_1\cdot\vec k_2}{k_1^2} =
  \frac{1}{2}\left(1+\tau_1\right)^2\;,
\label{eq:03-71}
\end{equation}
where the latter step follows because one of the remaining delta distributions ensures $\vec k_1=-\vec k_2$. Combining results, we find
\begin{equation}
  \bar Z_{123}^{(1)} = \frac{\bar\rho^3}{2}
  (2\pi)^6\delta_\mathrm{D}\left(\vec k_1+\vec k_2\right)
  \delta_\mathrm{D}\left(\vec k_3\right)
  \left(1+\tau_1\right)^2\,P_\delta\left(\vec k_1\right)\;.
\label{eq:03-72}
\end{equation}
The index combination $(j_1,j_2,j_3) = (2,1,3)$ adds the same expression. Taking the remaining cyclic index permutations into account leads to the bispectrum contribution
\begin{align}
  &G_{\rho\rho\rho}^{(1)}(123) = \bar\rho^3
  (2\pi)^6\left(1+\tau_1\right)^2 \nonumber\\ &\times
  \left\{
    \delta_\mathrm{D}\left(\vec k_1+\vec k_2\right)
    \delta_\mathrm{D}\left(\vec k_3\right)
    P_\delta\left(\vec k_1\right) + \mbox{cyc. perm.}
  \right\}
\label{eq:03-73}
\end{align}
for the bispectrum contribution from linear momentum correlations.

The terms of second order in the momentum correlation can be read off (\ref{eq:03-34}) and (\ref{eq:03-35}). The two-point term in $\bar Z_0^{(2)}[\tens L]$ gives
\begin{align}
  &\bar Z_{1212}^{(2)} = V^{N-3}(2\pi)^6
  \delta_\mathrm{D}\left(\vec k_1+\vec k_2\right)
  \delta_\mathrm{D}\left(\vec k_3\right)\tau_1^4 \nonumber\\ &\times
  \int\frac{\d^3k}{(2\pi)^3}P_\delta\left(\vec k\right)
  P_\delta\left(\vec k-\vec k_1\right)
  \left(\frac{\vec k_1\cdot\vec k}{k^2}\right)
  \left(\frac{\vec k_1\cdot(\vec k-\vec k_1)}
             {(\vec k-\vec k_1)^2}\right)\;.
\label{eq:03-74}
\end{align}
Since there are again four equivalent index configurations for this term, and since the three index combinations $(1,2)$, $(1,3)$ and $(2,3)$ are possible for the bispectrum, we arrive at the contribution
\begin{equation}
  G_{\rho\rho\rho}^{(2A)}(123) =
  \bar\rho(2\pi)^3\left\{
    \delta_\mathrm{D}\left(\vec k_3\right)G_{\rho\rho}^{(2)}(12) +
    \mbox{cyc. perm.}
  \right\}
\label{eq:03-75}
\end{equation}
of quadratic momentum correlations to the bispectrum, with $G_{\rho\rho}^{(2)}(12)$ taken from (\ref{eq:03-48}).

Finally, the three-point term $\bar Z_{jkkl}^{(2)}$ from (\ref{eq:03-35}) gives
\begin{align}
  \bar Z_{1223}^{(2)} &=
  V^{N-3}(2\pi)^3\delta_\mathrm{D}\left(\vec k_1+\vec k_2+\vec k_3\right)
  \tau_1^4
  \nonumber\\ &\times
  P_\delta\left(\vec k_1\right)P_\delta\left(\vec k_3\right)
  \frac{\vec k_1\cdot\vec k_2}{k_1^2}
  \frac{\vec k_2\cdot\vec k_3}{k_3^2}\;.
\label{eq:03-76}
\end{align}
Due to the symmetries of the $\bar Z_{jklm}^{(2)}$ terms, there are $8$ equivalent index combinations. This term thus contributes
\begin{align}
  &G_{\rho\rho\rho}^{(2B)}(123) = \e^{-Q_D/2}\bar\rho^3(2\pi)^3
  \delta_\mathrm{D}\left(\vec k_1+\vec k_2+\vec k_3\right)\tau_1^4
  \nonumber\\ &\times \left\{
    P_\delta\left(\vec k_1\right)P_\delta\left(\vec k_3\right)
    \frac{\vec k_1\cdot\vec k_2}{k_1^2}
    \frac{\vec k_2\cdot\vec k_3}{k_3^2} + \mbox{cyc. perm.}
  \right\}
\label{eq:03-77}
\end{align} 
to the bispectrum. Again, the damping factor should be approximated by (\ref{eq:03-68}).

Of the terms $G_{\rho\rho\rho}^{(1)}(123)$, $G_{\rho\rho\rho}^{(2A)}(123)$ and $G_{\rho\rho\rho}^{(2B)}(123)$ in (\ref{eq:03-73}), (\ref{eq:03-75}) and (\ref{eq:03-77}), only $G_{\rho\rho\rho}^{(2B)}(123)$ is a connected contribution to the bispectrum. Interchanging $\vec k_2$ and $\vec k_3$ there, and taking into account that the preceding delta distribution ensures $\vec k_3 = -(\vec k_1+\vec k_2)$, we can write
\begin{align}
  G_{\rho\rho\rho}^{(2B)}(123) &= \e^{-Q_D/2}\bar\rho^3(2\pi)^3
  \delta_\mathrm{D}\left(\vec k_1+\vec k_2+\vec k_3\right)\tau_1^4
  \nonumber\\ &\times
  \left\{
    P_\delta\left(\vec k_1\right)P_\delta\left(\vec k_2\right)
    F\left(\vec k_1,\vec k_2\right) + \mbox{cyc. perm.}
  \right\}
\label{eq:03-78}
\end{align}
with the kernel $F$ defined by
\begin{equation}
  F\left(\vec k_1,\vec k_2\right) := 1+
  \frac{\vec k_1\cdot\vec k_2}{k_1k_2}
  \left(\frac{k_1}{k_2}+\frac{k_2}{k_1}\right)+
  \frac{(\vec k_1\cdot\vec k_2)^2}{k_1^2k_2^2}\;.
\label{eq:03-79}
\end{equation} 
This is almost, but not quite, the $2F_2$-kernel appearing in Eulerian perturbation theory of the density contrast,
\begin{equation}
  2F_2\left(\vec k_1,\vec k_2\right) =
  \frac{10}{7}+\frac{\vec k_1\cdot\vec k_2}{k_1k_2}
  \left(\frac{k_1}{k_2}+\frac{k_2}{k_1}\right)+
  \frac{4}{7}\frac{(\vec k_1\cdot\vec k_2)^2}{k_1^2k_2^2}\;;
\label{eq:03-79a}
\end{equation} 
(cf.\ Eq.~(43) in \cite{2002PhR...367....1B}). The difference is due to our using the Zel'dovich approximation, which incorporates part of the non-linear structure growth.

\section{Summary}

This study is a first application of the non-equilibrium field theory for classical microscopic particles to cosmological structure formation. Based on the general formalism of the theory and on a free generating functional for initially correlated, canonical particle ensembles, we have derived two- and three-point cumulants of the cosmic density field. Our main results are:

\begin{itemize}
  \item If momentum correlations are taken into account to linear order, and if the free particle propagator is taken from the Zel'dovich approximation, the density power spectrum (\ref{eq:03-44}) reproduces the linear growth well known from standard perturbation theory.
  \item Evolving quadratic momentum correlations with the free Zel'dovich propagator leads to a first contribution to the nonlinear evolution of the power spectrum, for which the simple, closed expression (\ref{eq:03-69}) can be given. This contribution is a convolution of the initial power spectrum with itself, multiplied by a mode-coupling kernel.
  \item Deriving the bispectrum, we get the three terms (\ref{eq:03-73}), (\ref{eq:03-75}) and (\ref{eq:03-77}), only the third of which is connected. It can easily be brought into a form resembling the bispectrum result from Eulerian perturbation theory, but with a small difference in two coefficients.
\end{itemize}

These results have been obtained without invoking an interaction potential. Mildly nonlinear growth is possible nontheless because the Zel'dovich appoximation partially accounts for the interaction. Perturbation theory in the interaction potential to first and second order will be the subject of separate studies.

\begin{acknowledgments}
We wish to thank Luca Amendola, J\"urgen Berges, Marc Kamionkowski, Manfred Salmhofer, Bj\"orn Sch\"afer and Christof Wetterich for inspiring and helpful discussions. This work was supported in part by the Transregional Collaborative Research Centre TR~33, ``The Dark Universe'', of the German Science Foundation (DFG) and by the Munich Institute for Astro- and Particle Physics (MIAPP) of the DFG cluster of excellence ``Origin and Structure of the Universe''.
\end{acknowledgments}

\bibliography{../bibliography/main}

\end{document}